\documentclass[twocolumn,floatfix,prl,showpacs,superscriptaddress]{revtex4}

\usepackage{graphicx}
\usepackage{color}
\usepackage{bm}
\newcommand{\ket}[1]{| #1 \rangle}

\newcommand{\beq}{\begin{eqnarray}}
\newcommand{\eeq}{\end{eqnarray}}

\begin{document}
\title{Coupling Nitrogen-Vacancy Centers in Diamond to Superconducting Flux Qubits}

\author{D. Marcos}
\affiliation{
  Departamento de Teor\'ia de la Materia Condensada,
  Instituto de Ciencia de Materiales de Madrid, CSIC,
  Cantoblanco 28049, Madrid, Spain}
\author{M. Wubs}
\affiliation{QUANTOP, The Niels Bohr Institute, University of Copenhagen, 2100 Copenhagen \O, Denmark}
\affiliation{DTU Fotonik, Danish Technical University, 2800 Kgs. Lyngby, Denmark}
\author{J. M. Taylor}
\affiliation{Joint Quantum Institute and the National Institute of Standards and Technology, College Park, MD 20742}
\author{R. Aguado}
\affiliation{
  Departamento de Teor\'ia de la Materia Condensada,
  Instituto de Ciencia de Materiales de Madrid, CSIC,
  Cantoblanco 28049, Madrid, Spain}
\author{M. D. Lukin}
\affiliation{Department of Physics, Harvard University, Cambridge MA
02138}
\author{A. S. S\o rensen}
\affiliation{QUANTOP, The Niels Bohr Institute, University of Copenhagen, 2100 Copenhagen \O, Denmark}
\begin{abstract}
We propose a method to achieve coherent coupling between Nitrogen-vacancy (NV) centers  in diamond and superconducting (SC) flux qubits. The resulting coupling can be used  to create a coherent interaction between the spin states of distant NV centers mediated by the flux qubit. Furthermore, 
the magnetic coupling can be used to achieve a coherent transfer of quantum information between the flux qubit and an ensemble of NV centers. This 
enables 
a long-term memory for a SC quantum processor and possibly an interface between SC qubits and light.

\end{abstract}
 \maketitle

Among the many different approaches to quantum information processing, each
has its own distinct advantages. For instance, atomic systems~\cite{Cirac95} present
excellent isolation from their environment, and can be interfaced with 
optical photons for quantum communication. In contrast, condensed 
matter systems~\cite{Kane98Loss98} offer strong interactions, 
and may benefit from the stability, robustness,
and scalability associated with modern solid-state engineering.
In the last years, much effort is being devoted to coupling
atomic and solid-state qubits to form hybrid systems, combining 
``the best of two worlds".
One attractive approach to hybrid systems involves the integration of atomic ensembles 
with superconducting (SC) stripline resonators.  Strong coupling between SC qubits
and such resonators has already been achieved~\cite{Wallraff04}, and
approaches to extend the coupling to atomic systems
have been proposed~\cite{Tian04Sorensen04,Rabl06}. To achieve an
appreciable coupling, these proposals often use an electric
interaction, but magnetic interactions are more desirable, since
long coherence times are mainly achieved in systems where spin
states are used to store the information. Magnetic interactions are, however, inherently weaker but recently it has been proposed theoretically \cite{Rabl06,Imamoglu09Molmer09Schoelkopf09} and shown experimentally \cite{Steve10Schoelkopf10} that strong coupling to ensembles of spin systems can be achieved. 

In this Letter, we propose a novel hybrid system, consisting of a SC flux qubit magnetically coupled to Nitrogen-vacancy centers (NVs) in diamond. The
latter system shares many of the desirable properties of atoms, such as
extremely long coherence times and narrow-band optical transitions~\cite{Jelezko06}, but at the same time, 
the integration with solid state systems
can be relatively easy, as it eliminates the need for complicated trapping procedures. Additionally, much higher densities can be achieved with very limited decoherence~\cite{Imamoglu09Molmer09Schoelkopf09,Taylor08}. 
As we show below, the magnetic coupling between a 
SC flux qubit and a single NV center can be about three orders of magnitude stronger than that associated with stripline resonators, thereby making the system an attractive building block for quantum information processing.

\begin{figure}[t]
 \begin{center}
\includegraphics[width=3.0in]{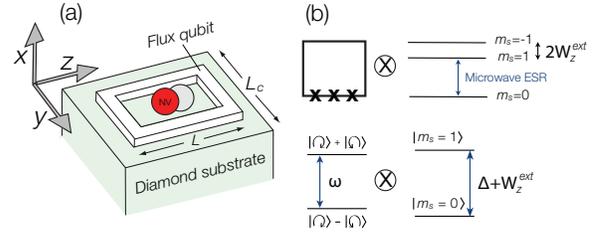}
 \caption{
 (a) Schematic setup. An isotopically pure ${}^{12}\rm{C}$ diamond crystal, doped with Nitrogen-vacancy color centers, is located in the proximity of a flux qubit of size $L \times L $ and cross section $h\times h$. The two persistent-current quantum states of the flux qubit have different associated magnetic fields, which give rise to a state-dependent interaction with the electron spin in the NV center(s). (b) The combined system of a flux qubit and a NV center. The eigenstates of the flux qubit are superpositions of left- and right-circulating currents. An external magnetic field splits the $m_S=\pm1$ states of the NV, resulting in a two-level system with the $m_S=0 \leftrightarrow m_S=1$ transition close to resonance with the flux qubit.
   \label{FQsketch}
}
 \end{center}
\end{figure}

Flux qubits (FQs) form superpositions of persistent currents of
hundreds of nano-Amperes, flowing clockwise and anti-clockwise through
micrometer-sized 
SC loops~\cite{Mooij99Wal00}. The
magnetic field associated with this current, of the order of a $\mu$T, enables a magnetic dipole coupling to the electron spins associated with crystalline impurities such as the NV
center in diamond.  Of particular interest is the coincidence of
energy splittings: the two states of the 
FQ are typically separated by a few GHz, while NV centers have a $S=1$ ground
state, with zero-field splitting $\Delta = 2 \pi \times 2.87$ GHz between
the $m_S=0$ and $m_S=\pm 1$ states.
By the application of a mT magnetic field, one of the spin transitions of the NV center can be tuned into resonance
with the FQ (see Fig.~\ref{FQsketch}(b)). 
This, together with the large magnetic moment of the FQ and the 
relatively long coherence times of both systems, opens the possibility 
of achieving coherent transfer between them.

Let us consider a {\em single} NV center in a diamond crystal, located
near a square FQ of size $L$ and thickness $h$ (see
Fig.~\ref{FQsketch}(a)). A static external field $\vec{B}^{\rm ext}$ is
applied, whose component perpendicular to the 
FQ provides half a flux quantum, and brings the qubit
near the degeneracy of the clockwise and counter-clockwise current states.  The zero-field spin splitting of the NV center $\Delta$
sets a preferred axis of quantization to be along the axis between the
Nitrogen and the vacancy; thus, the field parallel to this axis sets
the small additional Zeeman splitting between $m_S=\pm 1$ states, and
allows to isolate a two-level subsystem comprised by $m_S=0,1$.

For convenience, we denote as the $z$ axis the crystalline axis of the
NV center. The Pauli operators for the FQ system, not
tied to a particular spatial axis, will be denoted by
$\hat{\tau}_{1},\hat{\tau}_{2},\hat{\tau}_{3}$, with $\hat{\tau}_3$ describing the population difference between the two persistent-current states. 
The interaction of the total magnetic field $\vec{B}$
(external and from the FQ) with the NV center can be written 
as $\vec{S}\cdot \vec{W}$, where $ \vec{W} \equiv g_{e} \mu_B \vec{B}$, 
$g_{e}$ is the electron g-factor and $\mu_B$ is the Bohr magneton; the two
persistent-current quantum states of the FQ give rise to different
anti-aligned magnetic fields: $\hat{\tau}_3 \vec{W}^{\rm FQ}$.
The Hamiltonian for the system is then
\begin{equation}
  \hat{H} = \varepsilon \hat{\tau}_3/2 + \lambda \hat{\tau}_1+ \Delta S_z^2 + W^{\rm ext}_z S_z +
  \hat{\tau}_3 \vec{W}^{\rm FQ} \cdot \vec{S}.
  \label{eq:Hbare}
\end{equation}
Here the magnitude of $\vec{W}^{\rm FQ}$ corresponds to a Larmor frequency shift due
to the circulating or counter-circulating currents in the 
FQ, $\lambda$ is the coupling between these two current states, and
$\varepsilon$ is the bias in the two-well limit of the FQ, which
depends on the external field perpendicular to the loop. If the flux
qubit's plane is not perpendicular to the $z$ axis of the NV center, both
systems can be tuned on resonance by changing independently the $z$
and, e.g., $x$ components of $\vec{B}^{\rm ext}$.

Due to the coupling $\lambda$ in equation~(\ref{eq:Hbare}), the eigenstates of the FQ Hamiltonian are not  left- and right- circulating current states. This means that there is a magnetic transition between the dressed states of the FQ which couples to the electronic spin of the NV.  To describe this, we rotate the FQ via a
unitary transformation by an angle $\cos \theta \equiv
\varepsilon / 2\omega$, giving two FQ dressed states with a transition frequency $\omega \equiv \sqrt{\varepsilon^2/4 + \lambda^2}$.  When $\Delta + W^{\rm ext}_z - \omega =
\delta$ is small, we can transform to a rotating frame and make the
rotating-wave approximation (RWA) to describe the near-resonance interaction between the NV and the FQ.  Neglecting the state
$m_S = -1$, due to the external field moving it far out of resonance
($| \delta | \ll |W^{\rm ext}_z|$), the effective Hamiltonian describing the dynamics is
\begin{equation}
  \hat{H}_{\rm RWA}  =   \frac{\delta}{2} \hat{\sigma}_z + \frac{\cos \theta}{2}  W^{\rm
    FQ}_z \hat{\tau}_3 \hat{\sigma}_z
    + \frac{\sin \theta}{\sqrt{2}} W^{\rm FQ}_\perp { \hat{\tau}_- \hat{\sigma}_+ + {\rm H.c.}},
\end{equation}
where $\bm{\hat{\sigma}}$ are Pauli operators describing   the NV $m_S=0,1$ electron-spin states, and $\vec\tau$ is in a rotated basis so that, e.g., $\hat{\tau}_3$ describes the difference  in the populations of the FQ dressed states. 

\begin{figure}[t]
 \begin{center}
 \includegraphics[width=3.0in]{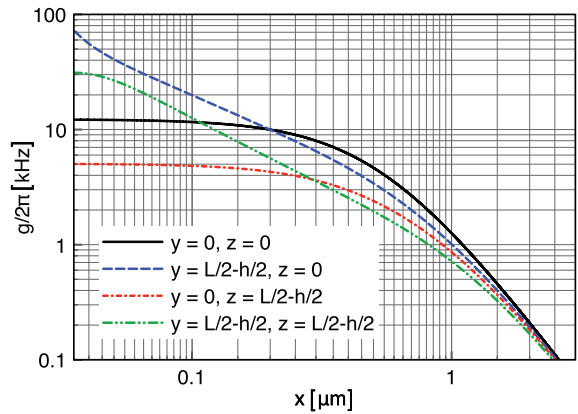}
 \caption{
 (Color online).
Coupling $g$ between a FQ operated at the degeneracy point ($\varepsilon=0$) and a single NV center located at the position $(x,y,z)$ in a reference frame with axes as in Fig.~\ref{FQsketch} but centered in the middle of the square FQ.
 The NV crystal axis is assumed to be parallel to one of the wires  forming the flux qubit.
 The FQ has size $L=1\,\mu\rm{m}$, thickness $h=60\,\rm{nm}$, and  a critical current of $0.5\,\mu \rm{A}$.
 \label{coupling1nv}
 }
 \end{center}
\end{figure}

The coupling constant $g \equiv \sin \theta W^{\rm
  FQ}_{\perp}/\sqrt{2}$ depends on the field perpendicular to
the NV axis, and it is maximal at the degeneracy point $\varepsilon=0$, where $\theta = \pi/2$.
Furthermore, this point has a ``sweet-spot" property: 
the energetics of the system is insensitive to small fluctuations of $\varepsilon$,
as the eigenvalues have zero derivative with respect to $\varepsilon$
at this point. This reduces the dephasing of the FQ due to stray magnetic fields and, e.g., paramagnetic spins in the diamond crystal \cite{AuxMat}, and ensures that there will be no differential shifts of the resonance frequency of the NVs due to an inhomogeneous static field from the FQ. For the remainder of this Letter we will assume that we are working at the degeneracy point.  Figure~\ref{coupling1nv} shows the coupling $g$, 
along four different vertical lines. We evaluate this using the magnetic field generated by a finite width square loop as given by the Biot-Savart law. 
For a FQ of size $L=1\,\mu\rm{m}$ and critical current $0.5\,\mu\rm{A}$, the
coupling reaches $g = 2 \pi \times 12\,\rm{kHz}$ for a single NV
center located at the center of the loop, which is about a factor
of 1000 larger than the coupling achieved with stripline resonators~\cite{Imamoglu09Molmer09Schoelkopf09}.  This coupling $g$ is however too small
to achieve coherent transfer,
since current $T_{2}$ times in FQs are at best 
 a few microseconds~\cite{FQcohRefs}.

We now show that the FQ can be used as a virtual intermediary, allowing to couple coherently two or more NV centers with the same orientation
and detuned $\delta$ from the FQ.
For large enough detunings, $\delta \gg 1/T_{2}^{{\rm FQ}}, g$, we can
adiabatically eliminate the excited state of the FQ, and the two
coupled NVs have the states $\ket{11}$ and $(\ket{10} +
\ket{01})/\sqrt{2}$ shifted by an energy $2 g^2/\delta$.
In contrast, the states $(\ket{10} -
\ket{01})/\sqrt{2}$ and $\ket{00}$ are not shifted by the FQ. As a result, for a fixed time $t_X = \pi \delta / (4 g^2)$ an operation
resembling $\sqrt{\textrm{SWAP}}$ -- an entangling operation -- between two NV centers can  be implemented.

To analyze the gate fidelity, we notice that the coherence of FQs and NVs are subject to low-frequency noise.
Larger coherence times are typically obtained by removing this noise during single-qubit evolution, using spin-echo sequences.  Such sequences can be incorporated into the gate operation by following the prescription for composite pulses of the
CORPSE family~\cite{Cummins03}. ÊIn particular, turning on and off the
interaction with the FQ, e.g., by changing from a $m_S = +1,0$
to $m_S=-1,0$ superposition, local $\pi$ pulses (described by $\hat{\sigma}_x$)
allow to realize a sequence $\textrm{Ex}_{\pi/4- \phi/2} \hat{\sigma}_x^1
\hat{\sigma}_x^2 \textrm{w}\textrm{Ex}_{\phi} \textrm{w}
\hat{\sigma}_x^1 \hat{\sigma}_x^2 \textrm{Ex}_{\pi/4-\phi/2}$, where
$\phi = 2 \sin^{-1} (1/\sqrt{8}) \approx 41.4^{\circ}$,
$\textrm{Ex}_{\theta}$ is an exchange-type interaction for a rotation
$\theta$, with $\theta = \pi$ a ``SWAP'', and $\textrm{w}$ is a wait
for a time $t_W = t_X (1/2 - 2 \phi/\pi)$, which makes the time
between the $\pi$ pulses equal to twice the time on either end of the
sequence. ÊThis approach is only sensitive to detuning errors in
fourth order for both collective and individual noise on the two
spins, effectively integrating a Carr-Purcell-type spin echo with the
$\sqrt{\textrm{SWAP}}$ operation. Furthermore, the total time
$t_X$ spent interacting with the FQ during the sequence is
equal to the prior case, thus inducing no additional overhead in the virtual coupling through the FQ.

With a spin echo the coherence is often limited by energy relaxation, resulting in an exponential decay $\exp(-t/T_2^{{\rm FQ}})$ for the FQ \cite{FQcohRefs}, whereas the NVs decay as $\exp(-(t/T_2^{{\rm NV}})^3)$ \cite{Taylor08,Maze08}. During the operation, the finite chance of exciting the detuned FQ leads to an
induced decoherence rate $\gamma = 2 g^2 / (
\delta^2 T_{2}^{{\rm FQ}})$, resulting in a gate error  $\sim  ((t_X+2t_W) / T_{2}^{{\rmÊNV}})^3 + \gamma t_X$. 
Minimizing this expression we find an optimal detuning, which gives the
optimized error probability $\sim 2.2/(g ^2T_2^{{\rm FQ}}
T_2^{{\rm NV}})^{3/4}$ \cite{AuxMat}. ÊFor isotopically purified $^{12}$C diamond \cite{Balasubramanian09} with an optimistic
$T_{2}^{{\rm NV}} \approx 20$~ms, $g \approx 2 \pi \times 12$~kHz, and
$T_{2}^{{\rm FQ}} \approx 5\ \mu$s, the maximum achievable fidelity of
the $\sim \sqrt{\textrm{SWAP}}$ operation is $\gtrsim 0.98$ with an operating time $t_X+2t_W\approx 3.3$ ms. It is thus possible to achieve a high-fidelity coherent operation between two NVs separated by micrometer distances. This coupling may even be extended to NVs separated by large distances: If two FQs are strongly coupled (directly or through resonators) with a coupling exceeding the detuning $\delta$, two NVs residing in different FQs may be coupled through the dressed states of the two FQs. This results in a long-distance coupling between the NVs of roughly the same magnitude as derived above. 

Even though the FQ coherence times are much shorter than the coupling to a single NV center, it is possible to coherently transfer the quantum-state  from the FQ to an ensemble of many ($N$) NVs by benefiting from a $\sqrt{N}$ enhancement in the coupling constant~\cite{Lukin03Hammerer09}.
Consider the diamond crystal depicted in Fig.~\ref{FQsketch}(a), with
density $n$ of NV centers, each of them with a fixed quantization axis
pointing along one of four possible crystallographic directions. If the orientations
are equally distributed among the four possibilities, and the external
field is homogeneous, a quarter of the
centers can be made resonant with the FQ. 
The total coupling will then be given by the FQ - single NV interaction summed over the resonant subensemble, whose state is conveniently expressed in terms of the collective angular momentum operator
$\hat{J}_{+} \equiv (1/G) \sum_{j} g_j
\hat{\sigma}_{+}^{(j)}$, where the sum runs over the resonant NVs,
and $G \equiv (\sum_{j} \vert g_j \vert^2)^{1/2}$
is the collective coupling constant. At the operating temperature of the FQ (tens of mK) the NVs are near full polarization, 
$\hat{\sigma}_z\approx\langle{\hat{\sigma}_{z}^{(j)}}\rangle \simeq -1$, and 
the collective spin operators fulfill
harmonic-oscillator commutation relations, thus having 
bosonic excitations in this limit.  
Tuning the system into resonance, the interaction between 
FQ and ensemble of NV centers is
 \beq
\hat{H}_{\rm{int}} = G \; \hat{\tau}_{-}\hat{J}_{+} +
{\rm H.c.} \label{Hcollective}, \eeq
whose dynamics can be complicated in general~\cite{Lopez07, Christ07},
but close to full polarization takes place between the states
$|{1}\rangle_{\rm{FQ}}|0^N\rangle_{\rm{NV}}$ and
$|{0}\rangle_{\rm{FQ}}\hat{J}_+|0^N\rangle_{\rm{NV}}$, where $|0^N\rangle_{\rm{NV}}$ corresponds to all NVs being in the ground state~\cite{Taylor03}. One can thus reversibly transfer the quantum state between the FQ and the collective excitations of the NVs. 

\begin{figure}[t]
 \begin{center}
 \includegraphics[width=3.0in]{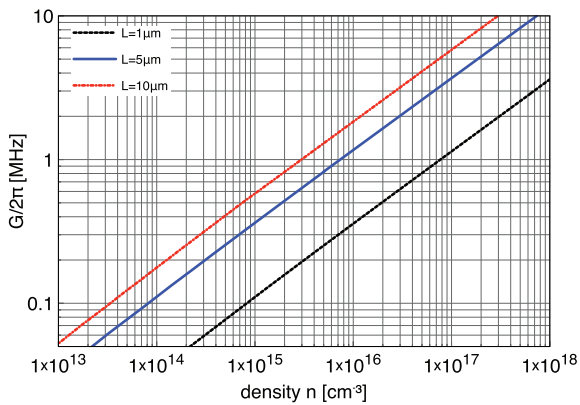}
 \caption{ (Color online). Coupling between a FQ and an ensemble of NV centers, as a function of the NV density $n$. Results are shown for three sizes $L$ of the square flux qubit, and are obtained by summing the inhomogeneous coupling constant, calculated as in Fig.~\ref{coupling1nv}, over an ensemble of NVs located in a cubic diamond crystal of size $L_C=2 L$ placed at a height $h/2$ below the FQ. In all cases, the NV crystal axis is assumed to be parallel to one of the wires forming the FQ, the width of the FQ is $h=60\,\rm{nm}$, and its critical current is  $0.5\,\mu\rm{A}$.
   \label{Gcoll}
}
 \end{center}
\end{figure}

The collective coupling $G$ as a function of the density $n$ of NV centers, 
is shown in Fig.~\ref{Gcoll} for three different sized FQs. The coupling is obtained by summing the inhomogeneous coupling over NVs distributed in a crystal of size $(2L)^3$.
Taking the FQ coherence time to be $ \sim 5\,\mu\rm{s}$, 
we find that for a FQ with $L =5\,\mu\rm{m}$, coherent
transfer becomes possible at densities $n \gtrsim
10^{16}\,\rm{cm}^{-3}$, which have already been achieved in recent experiments~\cite{Waldermann07}. 
We note that it might be possible to increase the interaction strength with the FQ by designing circuits with higher critical currents.
Furthermore, an important feature of the present approach is that strong coupling to the FQ is achieved with ensembles containing a relatively small number of spins in a few micrometers, which makes it easier to achieve fast and identical manipulation of all spins, e.g., in spin-echo approaches.

The performance of the transfer of information between FQ and collective ensemble will be limited by paramagnetic impurities present in the diamond crystal, which will interact via a dipolar coupling with the NVs encoding the collective quantum state. In the Supplementary Information we give a detailed discussion of decoherence induced by these impurities, which, due to the typically low Nitrogen to NV conversion efficiency, is dominated by unpaired Nitrogen electrons in the sample \cite{AuxMat}. For a FQ of size $L=5\ \mu$m and a density of $n=10^{17}$ cm$^{-3}$ we estimate a coherence time of $T_2^*=1.8\ \mu$s. This is sufficiently long for the infidelity induced by the paramagnetic impurities to be on the percent level, both for the transfer from the FQ to the spin wave and the nuclear storage discussed below. Furthermore the decoherence of the FQ induced by the impurities is negligible if we work close to the degeneracy point of the FQ, $\cos\theta\approx 0$.

So far we have ignored the influence of the nuclear spin.
The strong hyperfine interaction with lattice $^{13}$C will be detrimental to the presented schemes, but this can be overcome using isotopically purified $^{12}$C diamond \cite{Balasubramanian09}. For the Nitrogen atoms forming the centers there are no stable isotopes without nuclear spin. This spin can, however, be polarized by transferring the nuclear state to the electron spin,
using a combination of radio-frequency and microwave pulses,
followed by polarization of the electron spin~\cite{Childress06Gurudev07}, 
or directly through optical pumping using excited-state couplings~\cite{Childress09}. For $^{14}$N, the  large quadrapolar field ($\sim$ 5 MHz) from the NV center leads to quantization of the spin-1 nucleus  along the NV axis. This suppresses spin-flip terms such that the hyperfine interaction just acts as an
additional parallel component to the external field.
Due to the long ($\gg 1$ s) relaxation
time of the $^{14}$N nuclear spin, this can be accounted for by an
appropriate detuning of the magnetic field so that the nuclear spin does not lead to decoherence.  
  
The nuclear spin can actually be turned into a valuable resource for long-term storage of quantum information. The electronic spin state ($m_S$) can be  transferred into the nuclear spin state ($m_I$) through a sequence of radio and microwave frequency pulses performing the evolution $\alpha\ket{00}+\beta\ket{10}\rightarrow \alpha \ket{00}+\beta\ket{11} \rightarrow \alpha \ket{00} + \beta\ket{01}$, where the states label the magnetic quantum numbers $\ket{m_S m_I}$.
This allows for a long-term memory in the system while other operations are performed, e.g., while the NVs interact with light for quantum communication.

We have shown how to magnetically couple a superconducting FQ to NV centers in diamond. This may be used to achieve strong coupling of distant centers or to transfer the state of a FQ to an ensemble of NVs. The latter could be used for long-term nuclear storage, and, using the strong optical transitions of the NVs, may enable an interface between superconducting qubits and light. 

We thank P. Forn-D\'iaz, K. M\o lmer, R. Schoelkopf, A. Imamo\u glu, K. Semba and J. E. Mooij for discussions. This work was supported by NSF, CUA, DARPA and Packard Foundation, and grants FPU AP2005-0720, FTP 274-07-0080, FIS2009-08744 and CCG08-CSIC/MAT- 3775.
At the time of submission a related preprint appeared \cite{Barrett09},
which describes an NV  coupled to a persistent-current loop
integrated into coplanar resonators.

\end{document}


\title{Coupling Nitrogen Vacancy Centers in Diamond to Superconducting Flux Qubits: Supplementary Information}

\author{D. Marcos}
\affiliation{
  Departamento de Teor\'ia de la Materia Condensada,
  Instituto de Ciencia de Materiales de Madrid, CSIC,
  Cantoblanco 28049, Madrid, Spain}
\author{M. Wubs}
\affiliation{QUANTOP, The Niels Bohr Institute, University of Copenhagen, 2100 Copenhagen \O, Denmark}
\affiliation{DTU Fotonik, Danish Technical University, 2800 Kgs. Lyngby, Denmark}
\author{J. M. Taylor}
\affiliation{Joint Quantum Institute and the National Institute of Standards and Technology, College Park, MD 20742}
\author{R. Aguado}
\affiliation{
  Departamento de Teor\'ia de la Materia Condensada,
  Instituto de Ciencia de Materiales de Madrid, CSIC,
  Cantoblanco 28049, Madrid, Spain}
\author{M. D. Lukin}
\affiliation{Department of Physics, Harvard University, Cambridge MA
02138}
\author{A. S. S\o rensen}
\affiliation{QUANTOP, The Niels Bohr Institute, University of Copenhagen, 2100 Copenhagen \O, Denmark}

\begin{abstract}
In the main text we have omitted several details from the estimates of decoherence and the performance of the procedures. In the following we give more details on how these were obtained.
\end{abstract}
 \maketitle

\section{Decoherence of the NV centers}

\subsection{$\sqrt{{\rm SWAP}}$ operation between individual NV centers}

In the main text, we describe a method to perform an entangling
operation between two NV centers near the same FQ using this as a quantum bus. 
The interaction between different NVs is mediated off-resonantly by the FQ and can be described starting from the Hamiltonian $\hat{H}_{\rm RWA}$ in the main text, extended to the case of $2$ NVs. 
Taking the mixing angle $\theta
\rightarrow \pi/2$, such that the FQ is optimally biased with respect
to low-frequency flux noise, and a rotating frame such that the FQ is detuned $-\delta$, rather
than the NVs being detuned $\delta$, this Hamiltonian reads
\begin{equation}
H = -\frac{\delta}{2}( \hat{\tau}_3 +1)+ \frac{1}{\sqrt{2}} \hat{\tau}_- \left(
W_{\perp,1}^{FQ}  \hat{\sigma}_+^{(1)} + W_{\perp,2}^{FQ} \hat{\sigma}_+^{(2)} \right) + {\rm H.c.}
\end{equation}
In this picture, it becomes clear that there are two zero-energy
eigenstates: $H \ket{00}\ket{0}_{FQ} = H \ket{D} \ket{0}_{FQ} = 0$, with the
dark state
$\ket{D} \equiv \frac{1}{\sqrt{2} G_2} \left(
  W_{\perp,2}^{FQ} \ket{10} - W_{\perp,1}^{FQ} \ket{01} \right)$ and the collective coupling constant for the two NVs $G_2 = \sqrt{(W_{\perp,1}^{FQ})^2 + (W_{\perp,2}^{FQ})^2}/\sqrt{2}=\sqrt{g_1^2+g_2^2}$.  There is a corresponding bright state $\ket{B} = \frac{1}{\sqrt{2} G_2} \left(  W_{\perp,1}^{FQ} \ket{10} + W_{\perp,2}^{FQ} \ket{01} \right)$, and the state $\ket{B} \ket{0}_{FQ}$ is coupled to
$\ket{00}\ket{1}_{FQ}$ with coupling strength $G_2$. Similarly the state
$\ket{11}\ket{0}_{FQ}$ is coupled to $\ket{B} \ket{1}_{FQ}$ with coupling
strength $G_2$.

Therefore, in second order perturbation theory, projecting onto the lower
energy state of the flux qubit, we have 
\beq 
H_{\rm eff} =
\frac{G_2^2}{\delta} (\ketbrad{B} + \ketbrad{11})\ . 
\eeq 
Here we will assume $W_{\perp,1}^{FQ}\approx W_{\perp,2}^{FQ}$ and thus $G_2\approx \sqrt{2} g$. The
probability of finding the system in the excited state is then zero for the
two dark states and $2 g^2/\delta^2$ for the two bright states of the
system.  Thus, FQ relaxation and Markovian dephasing, which occur
directly to the FQ at a rate $1/T_2^{\rm FQ}$, will lead to errors at a rate
$\sim {2 g^2}/{\delta^2 T_2^{\rm FQ}}$ for the two NV spins. Low-
frequency noise enters in small variations of $\delta$, which also
contribute at order $2 g^2/\delta^2$. Since such low-frequency noise is non-Markovian it will only give a quadratic contribution  $\sim( g^2 t/(\delta^2 T_2^{\rm FQ}))^2 $, which will be less severe than the Markovian dephasing. We therefore take a worst-case scenario and assume the decoherence to be produced by Markovian dephasing.

To optimize the gate operation, we note that the overall time when the FQ is coupled is $t_{X} = \frac{\pi}{2} \frac{\delta}{2 g^2}$.  There is an additional wait time for the NV centers of $2t_W = t_X (1 - 4 \phi/\pi)$, where $\phi =2 \sin^{-1} (1/\sqrt{8})$ as in the main text.  For NV spins in the states $\ket{00}, \ket{D}$, the error is entirely from the total time $t_X + 2t_W$, as the FQ is not included, while for NV spins in the states $\ket{11},\ket{B}$, both FQ and NV dephasing errors enter.  We take the latter case to overestimate the error induced in the operation, and find an effective dephasing of the NV center $\exp[- (\frac{t_X + 2 t_W}{T_2^{\rm NV}})^3]$, as expected from a dipole-dipole bath decorrelation, and an additional dephasing due to the admixture of the FQ as $\exp[-\frac{2 t_X g^2}{\delta^2T_2^{\rm FQ}}]$.  The only free parameter we can optimize over is the detuning $\delta$.  Rewriting the total fidelity as 
\beq
F = \exp[- \frac{\delta^3}{\alpha^3} - \frac{\beta}{\delta}]
\eeq
with $\alpha = \frac{2g^2 T_2^{\rm NV}}{\pi(1 - 2 \phi/\pi)}$ and $\beta = \frac{\pi}{2 T_2^{\rm FQ}}$, the optimum occurs for $\delta_* = ( \alpha^3\beta/3)^{1/4}$, giving $F_* = \exp[ - (\beta/\alpha)^{3/4} (3^{1/4} + 3^{-3/4})]$.  Substituting $\alpha$ and $\beta$ in $\delta_*$ and the numbers given in the main text we find $F_* = \exp[ - 2.18 (g^2 T_2^{\rm FQ} T_2^{\rm NV})^{-3/4}] \approx 0.98$ and $\delta_* = 0.96 [ g^6 (T_2^{\rm NV})^3/T_2^{\rm FQ}]^{1/4} \approx 2 \pi \times 3.6$ MHz.

 An additional error arises when $W_{\perp,1}^{FQ} \neq W_{\perp,2}^{FQ}$, since then the unitary operation is not performing exactly the desired evolution; this leads to a reduced fidelity of the entangling operation by a factor $W_{\perp,1}^{FQ} W_{\perp,2}^{FQ}/g^2$. It is important to note, however, that this imperfection is not a decoherence effect and it may be possible to exploit the resulting unitary evolution even despite this imperfection, i.e., the resulting operation still resembles a $\sqrt{{\rm SWAP}}$.

\subsection{Decoherence of the ensemble of NV centers}

In the main article we described the transfer of an excitation from the FQ to the collective state of an ensemble of NV centers. 
In this section we estimate the dephasing of this collective state of the ensemble due to magnetic dipole-dipole interactions with other spins in the diamond crystal. The paramagnetic impurities in the diamond crystal consist of both other NVs centers with either the same or different orientation as well as unpaired electronic spins on substitutional Nitrogens. In most experiments the ratio of Nitrogen to NV centers is typically quite low, e.g. on the order of 1\% to 10\% \cite{Aharonovich09}. Below we refer to this ratio as the conversion efficiency. 

The dipole-dipole interaction is described by the Hamiltonian \cite{AbragamBook}
\begin{equation}
H_{{\rm dip}}=\sum_{j\neq k} \frac{1}{2} \frac{\mu_0}{4\pi r_{jk}^3}{\left[(\vec{m}^{(j)}\cdot \vec{m}^{(k)}-3(\vec{m}^{(j)}\cdot\vec{e}_{jk})( \vec{m}^{(k)}\cdot\vec{e}_{jk})\right]},
\end{equation}
where $r_{jk}$ is the distance between the dipoles $j$ and $k$, $\vec{e}_{jk}$ is a unit vector between them, and  $\vec{m}^{(j)}$ and $\vec{m}^{(k)}$ are their magnetic moments. These can be expressed in terms of the spin operators of the impurities, but the quantization axis for the different spins will be different: NV centers will be aligned along the symmetry axis of the center whereas electron spins on Nitrogens will be quantized along the axis of the applied static field. To describe this situation we introduce spin operators $\hat s_z^{(j)}$, $\hat s_+^{(j)}$ and $\hat s_-^{(j)}$, defined relative to the quantization axis of each particular spin, and take the different orientations into account in the coupling constant describing  the interaction between different spins. 

For the NV centers we furthermore ignore the $m_S=-1$ state which is assumed to be shifted out of resonance by a magnetic field. All the spins are therefore two-level systems which can be described by Pauli matrices $\hat{\sigma}_z^{(j)}$, $\hat{\sigma}_+^{(j)}$ and $\hat{\sigma}_-^{(j)}$. The spin operators  can then be expressed as 
\begin{equation}
\begin{split}
\hat s_z^{(j)}=\frac{1}{2}(\hat{\sigma}_z^{(j)}+l_j),
\qquad \hat s_+^{(j)}=\sqrt{1+l_j}\hat{\sigma}_+^{(j)},
\end{split}
\end{equation}
where the quantity $l_j$, which is unity for NV centers and vanishes for Nitrogen spin, accounts for the fact that the Nitrogen has spin 1/2 whereas the two-level NV system is made from the $m_S=0$ and $m_S=1$ states of a spin-1 particle. The magnetic dipole Hamilonian is then given by 
\begin{equation}
\begin{split}
H_{{\rm dip}}=& \sum_{j,k\neq j}
  a_{jk}\hat{\sigma}^{(j)}_{+}\hat{\sigma}^{(k)}_{-} +
  b_{jk} (\hat{\sigma}^{(j)}_{z}+l_j)(\hat{\sigma}^{(k)}_{z}+l_k),
\end{split}
\label{eq:Hd}
\end{equation}
where we have used the rotating-wave approximation to ignore terms which do not conserve energy. By this approximation, the first term vanishes between spins with different resonance frequencies, i.e., with different orientations. Notice, that if the quantization axes for spin $j$ and $k$ are different, then the angle between these axes can be contained in the coupling constants $a_{jk}$ and $b_{jk}$. This is the case for the interaction between the NV and Nitrogen spins, whose quantization axes, determined by the crystal axis and the external magnetic field respectively, form  an angle $\beta$ between them. For the sake of simplicity, we here only present the case $\beta=0$. This is a worst-case scenario for the $T_2^*$ derived below, but the general situation $\beta\neq 0$ only leads to minor modifications. 
With  parallel  quantization axes, we have
\begin{equation}\label{couplingconstants}
\begin{split}
a_{jk}&= \mu_{jk}^{(a)} \left[ 1- 6 e_{-}^{jk} e_{+}^{jk} \right] = \mu_{jk}^{(a)} \left[ 1 - \frac{3}{2} \sin^2\Theta_{jk} \right],\\
b_{jk}&= \mu_{jk}^{(b)} \left[ 1- 3 (e_z^{jk})^2 \right] = \mu_{jk}^{(b)} \left[ 1 - 3 \cos^2\Theta_{jk} \right],
\end{split}
\end{equation}
where $\Theta_{jk}$ is the angle between the vector $\vec{e}_{jk}$ and the crystal axis, $e_\pm^{jk}:=(e_x^{jk}\pm i e_y^{jk})/2$, and
\begin{equation}
\mu_{jk}^{(a)}:=\frac{1}{2}\frac{\mu_0}{4\pi} \frac{\mu_B^2 g_e^{(j)}g_e^{(k)} \sqrt{1+l_j}\sqrt{1+l_k}}{r_{jk}^3},\qquad \mu_{jk}^{(b)}:=\frac{1}{8}\frac{\mu_0}{4\pi} \frac{\mu_B^2 g_e^{(j)}g_e^{(k)} }{r_{jk}^3},
\end{equation}
with $\mu_B$ the Bohr magneton and $g_e^{(j)}$ the electron $g$-factor of the $j$th spin.

To evaluate the dephasing of the spin wave we will assume that we start out in a state $|00...0\ra_{{\rm NV}}$ where all the NVs are initially prepared in their ground states, and that a superposition state $c_0|0\ra_{{\rm FQ}}+c_1|1\ra_{{\rm FQ}}$ is transferred into the collective spin wave, resulting in a state $(c_0+c_1J_+)|00...0\ra_{{\rm NV}}$, with $J_+=(1/G)\sum_j g_j\hat\sigma_+^{(j)}$. We then evaluate the time evolution of the coherence
\begin{equation}
\la J_-(t)\ra={\rm Tr} {\left[ e^{iHt}J_{-}e^{-iHt} (c_0+c_1J_+)|00...0\ra_{{\rm NV}}\la 0 ...00|(c_0^*+c_1^*J_-) \rho_{{\rm B}}
\right]},
\end{equation}
where $\rho_{{\rm B}}$ is the initial density matrix of the bath. Since the Hamiltonian (\ref{eq:Hd}) conserves the number of excitations in the NVs,  this expression can be simplified to 
\begin{equation}
\begin{split}
\la J_-(t)\ra=c_0^*c_1
{\rm Tr} {\left[ e^{iHt}J_-e^{-iHt} J_+|00...0\ra_{{\rm NV}}\la 0 ...00| \rho_{{\rm B}}
\right]}.
\end{split}
\end{equation}

Writing out the above equation in terms of single-spin operators we see that the dephasing is determined by the time evolution of the two-point correlation function $\la \hat{\sigma}_-^{(j)}(t) \hat{\sigma}_+^{(k)}(t=0)\ra$ in the state $|00...0\ra_{{\rm NV}}\la 0 ...00| \rho_{{\rm B}}$. Since the density of Nitrogen is much higher than the density of NVs, the dephasing of the spin wave will predominantly be due to the interaction with the electronic spins of Nitrogen atoms, and we shall therefore ignore the interaction among the NVs. In this approximation there is no longer a mechanism in the Hamiltonian which can transfer the excitation from one NV to another and the correlation function $\la \hat{\sigma}_-^{(j)}(t) \hat{\sigma}_+^{(k)}(t=0)\ra$ vanishes exactly for $j\neq k$. The dephasing of the spin waves thus reduces to the calculation of the single-spin dephasing averaged over the spin wave:
\begin{equation}
\la J_-(t)\ra=c_0^*c_1 \sum_j \frac{|g_j|^2}{G^2}
{\rm Tr} {\left[ e^{iHt}\hat{\sigma}_-^{(j)}e^{-iHt} \hat{\sigma}_+^{(j)}|00...0\ra_{{\rm NV}}\la 0 ...00| \rho_{{\rm B}}
\right]},
\end{equation}
where the sum runs over all NVs in the spin wave. Assuming all NVs to be equivalent this can be simplified to the calculation of the dephasing of a single-spin in a bath
\begin{equation}
\la J_-(t)\ra=\la J_-(t=0)\ra {\rm Tr} {\left[ e^{iHt}\hat{\sigma}_-^{(j)}e^{-iHt} \hat{\sigma}_+^{(j)}|0\ra_{{\rm NV}}\la 0| \rho_{{\rm B}}
\right]}.
\label{eq:CohSingle}
\end{equation}
When considering a single NV, the action of the spin bath described by (\ref{eq:Hd}) essentially corresponds to a random magnetic field generated from the spin ensemble along the crystal axis. This effective field will be fluctuating in time because the dipole-dipole interaction among the Nitrogen spins introduces flip-flop processes.  When the Nitrogen concentration is much higher than that of NVs, the distance between Nitrogens is comparable to the distance between NVs and the nearest Nitrogen. These flip-flop processes will therefore take place on a time scale which is comparable or only slightly faster than the NV dephasing time. For simplicity we here  ignore the flip-flop processes and consider a static environment. This represents a worst-case scenario, since this approximation removes the time averaging of the field from these processes. The actual dephasing time $T_2^*$ will therefore be slightly larger than what we predict here. Within this approximation the expression in Eq. (\ref{eq:CohSingle}) can be reduced considerably:
\begin{equation}
\la J_-(t)\ra=\la J_-(t=0)\ra {\rm Tr} {\left[\prod_k e^{-i4b_{jk}\hat{\sigma}_z^{(k)}t}  \rho_{{\rm B}}
\right]},
\label{eq:CohSingleSimp}
\end{equation}
where the product is over the Nitrogen spins. 

With expression (\ref{eq:CohSingle}) we can evaluate the dephasing for a given bath. For a given density $n$, the typical strength of the spin-spin interaction is $\mu_0\mu_B^2 g_e^2 n/4\pi$, of the order of $\sim 25\;\mu$K for high Nitrogen densities $n_{{\rm N}}\sim 10^{19}$ cm$^{-3}$. This is much smaller than the  typical operating temperature of the FQ (tens of mK), and we can therefore neglect the interaction among the spins for determining the initial density matrix of the bath $\rho_{{\rm B}}$. Furthermore any non-vanishing mean value $\la \hat{\sigma}_k^{(j)}\ra\neq 0$ only results in a mean shift of the resonance frequency of the ensemble. The dephasing will thus be determined by spin fluctuations. A worst-case scenario can be obtained by assuming that the mean value vanishes $\la \hat{\sigma}_k^{(j)}\ra=0$, in which case the variance is maximal. The evolution of the coherence is then given by 
\begin{equation}
\la J_-(t)\ra=\la J_-(t=0)\ra \prod_k \cos(4b_{jk}t).
\end{equation}
To simplify this expression we expand it in time and find
\begin{equation}
\la J_-(t)\ra=\la J_-(t=0)\ra {\left(1- 2 t^2 n_{{\rm N}}\int d^3\vec r_k b_{jk}^2\right)},
\end{equation}
where we have replaced the sum by an integral and introduced the Nitrogen density $n_{{\rm N}}$. Using Eq. (\ref{couplingconstants}) this expression reduces to 
\begin{equation}
\begin{split}
\la J_-(t)\ra=\la J_-(t=0)\ra {\left[1- t^2
{\left(\frac{\mu_0}{4\pi} \frac{\mu_B^2g_e^{{\rm N}} g_e^{{\rm NV}}}{8}\right)}^2 4\pi n_{{\rm N}} \int_0^{\pi} \sin\Theta (1-3\cos^2\Theta)^2 d\Theta \int \frac{1}{r^4} dr \right]}.
\end{split}
\end{equation}
The angular integral gives $8/5$ but the radial integral has a strong divergence at $r\to 0$. Since the integral is an integral over the distance between the NV and the Nitrogen impurity, this divergency represents the very fast dephasing of NV centers which happen to have a nearby Nitrogen spin. Such NV centers with a nearby Nitrogen spin will, however, also be far out of resonance  during the interaction with the Flux qubit.   
This interaction takes a time $\sim 1/G$, and therefore any NV with a dipole interaction stronger than $G$ will effectively not participate in the spin wave. We exclude these NVs by truncating the integral at a distance $r_{\rm min}$ when the interaction strength reaches the value of the coupling constant $G$, i.e., 
\beq
G=\frac{1}{8}\frac{\mu_0}{4\pi} \frac{\mu_B^2 g_e^{N}g_e^{NV} }{r_{\rm min}^3}.
\eeq
This truncation of the integral gives
\begin{equation} \label{Jdecay}
\la J_-(t)\ra=\la J_-(t=0)\ra {\left[1- t^2 
{{\frac{\mu_0\mu_B^2g_e^{{\rm N}} g_e^{{\rm NV}}}{15\eta} n_{{\rm NV}} G}} \right ]},
\end{equation}
where $\eta$ is the Nitrogen-to-$NV$ conversion efficiency. The result (\ref{Jdecay}) corresponds to the first term in an expansion of an exponential decay $\exp(-(t/T_2^*)^2)$ with coherence time
\begin{equation}
T_2^*=\frac{1}{\sqrt{\frac{\mu_0\mu_B^2g_e^{{\rm N}} g_e^{{\rm NV}}}{15\eta} n_{{\rm NV}} G}}.
\end{equation}
Notice that in order to be applicable, our regularization procedure requires $r_{{\rm min}}$ to be much smaller than the typical distance between spins $1/\sqrt[3]{n_{{\rm N}}}$, since we require that only a small fraction of the NVs are excluded. This condition is fulfilled in the interesting regime $G T_2^*\gg1$. Taking a conversion efficiency $\eta=0.05$, we find $T_2^*\approx 0.3 \ \mu$s for a FQ with $L=5\ \mu$m and a density $n_{{\rm NV}}=10^{18}$ cm$^{-3}$, corresponding to $G\approx 2\pi\times 15$ MHz. For a full transfer of the 
state from the FQ to the spin wave and back we need a transfer time $t=\pi/G$, corresponding to an infidelity of the order of $1-F\sim(t/2T_2^*)^2 < 0.5\%$ (the factor of 2 accounting for the fact that the excitation only spends half of the transfer time in the spin wave). For different densities the infidelity scales as $1-F\propto \sqrt{n_{{\rm NV}}}$, becoming smaller at lower densities due to the reduced dipole-dipole interaction, e.g., at $n_{{\rm NV}}=10^{16}$ cm$^{-3}$ the error is reduced by an order of magnitude but then we are approaching the limit where the FQ decoherence becomes important. The transfer of excitations from the FQ to the spin wave is thus feasible in the regime $10^{16}$ cm$^{-3} \lesssim n_{{\rm NV}} \lesssim 10^{18}$ cm$^{-3}$. In order for the spin system to be useful as a long-term memory the coherence time should, however, also be sufficiently long to allow for the transfer  to the nuclear spin for long-term storage. Since this can at best be achieved on a time scale set by the hyperfine interaction ($\sim 5$ MHz), this would exclude working at the highest densities in this interval. Working at a density of $n_{{\rm NV}}\sim 10^{17}$ cm$^{-3}$ leads to $T_2^*\approx  1.8 \mu$s, which is sufficient to allow a transfer of the excitation from the electron spin to the nuclear spin with an infidelity $1-F$ at the percent level.  

The estimates above indicate that it is realistic to achieve a transfer of excitations from the FQ to the spin wave and back even without extending the coherence time by spin-echo techniques. These may, however, be desirable in order to achieve even longer coherence times. In particular $T_2$ may be extended if the NVs can be flipped by an external AC field on a time scale faster than the Nitrogen flip-flop processes. One should, however, be careful about applying spin-echo to the collective NV spin since, e.g. errors in the pulse area may give rise to collective decoherence processes. These can be more detrimental to quantum states stored in collective degrees of freedom than in individual spins (for instance flipping the $|0\ra-|1\ra$ transition would require control of the pulse area to an accuracy better than $1/\sqrt{N}$ in order to preserve the collective state). Since the dephasing of the NV spin wave is dominated by Nitrogen spins a more desirable solution could be to apply the external driving field to the electron spins on the Nitrogen atoms. If these spins are flipped on a time scale much faster than the flip-flop processes it would lead to an increased coherence time of the NV spin wave.  Since Nitrogen spins have a considerably different resonance frequency, this could be achieved with little influence on the NV spin wave where the quantum information is stored.

\section{Decoherence of the Flux qubit due the spin bath}
A different concern is the extent to which the diamond crystal with a high density of spins may cause dephasing of the flux qubit, of particular importance for the coupling with an ensemble of NVs. As discussed above, the paramagnetic impurities in the diamond crystal consist primarily of NV centers and unpaired electron spins on Nitrogen atoms which were not converted into NVs centers during the annealing process. Due to the typically low conversion efficiency from Nitrogen to NV centers, we shall first consider the effect of Nitrogen impurity spins and then discuss the role of the NV centers. The coupling of the FQ to paramagnetic impurities can be described by a Hamiltonian similar to Eq. (1) in the main text:
\begin{equation}  \hat{H} = \hat{H}_{{\rm spin}}+\varepsilon \hat{\tau}_3/2 + \lambda \hat{\tau}_1+ 
  \hat{\tau}_3 \sum_j \vec{W}^{\rm FQ}(\vec{r}_j) \cdot \vec{S}^{(j)},
  \label{eq:Hbare}
\end{equation}
where the spin Hamiltonian $\hat{H}_{{\rm spin}}=\sum_j \Delta_j S_z^{(j)}/2+\hat H_{{\rm int}}$ describes an energy splitting $\Delta_j$ of the individual spins and the interaction between them, encapsulated in $\hat H_{{\rm int}}$. Changing to the dressed-state picture of the flux qubit, this Hamiltonian is transformed into
\begin{equation}  \hat{H} = \hat{H}_{{\rm spin}}+\omega \hat{\tau}_3/2 + 
(\cos\theta  \hat{\tau}_3-\sin\theta\hat{\tau}_1) \sum_j \vec{W}^{\rm FQ} (\vec{r}_j)\cdot \vec{S}^{(j)}.
  \label{eq:Hdressed}
\end{equation}
In the rotating frame with respect to $\omega\hat{\tau}_3/2$, the operator $\hat{\tau}_3$ remains stationary whereas $\hat{\tau}_1$ oscillates at a frequency $\omega$ -- much higher than any time scale of the bath, provided that there are no near resonance impurities in the diamond sample. This is the case for the Nitrogen spins, whose splitting is determined by the applied field. The NVs are near resonance and will be dealt with below. The slowly varying and rapidly oscillating terms have different qualitative behavior so we will consider them separately. 

Let us first consider the slowly-varying contribution. As we will show now this contribution vanishes if we work close to the degeneracy point of the FQ $\cos\theta\approx 0$ where the left- and right-circulating current states are degenerate. The slowly-varying contribution is described by
\beq
H_\mathrm{slow} = \hat{\tau}_3 W_\mathrm{eff}^{FQ},
\eeq
with
\beq
W_\mathrm{eff}^{FQ} \equiv cos(\theta) \sum_j \left[ W_z^{FQ}(\vec{r}_j) \cos(\beta_j) - W_{\perp}^{FQ}(\vec{r}_j) \sin(\beta_j) \right] S_3^{(j)},
\eeq
and  $\beta_j$ being the angle between the quantization axis the $j$th spin and the crystal axis of the NV (recall that the $z$-axis is defined by the crystal axis). For simplicity, we define
\beq
\kappa_j := W_z^{FQ}(\vec{r}_j) \cos(\beta_j) - W_{\perp}^{FQ}(\vec{r}_j) \sin(\beta_j).
\eeq
Since the interaction of the flux qubit with an individual spin is on the order of $g$, the former has little influence on the state of an individual spin for the duration of the interaction $\sim 1/G\sim 1/g\sqrt{N}$. Furthermore the coupling of the FQ to collective degrees of freedom in the spin bath, will only have a limited influence  on the state of an unpolarized bath, since the FQ can flip at most a single-spin. We can therefore ignore the influence of the FQ on the spin bath and consider $W_\mathrm{eff}^{FQ}$ as a fluctuating external field. Any mean value of this field will merely give rise to a shift of the resonance frequency which can be compensated and we thus need to consider the fluctuations. The root mean square of the fluctuating field is given by 
\beq
\delta W_\mathrm{eff}^{FQ} &=& \cos(\theta) \Big[ \sum_{j,k} \kappa_j\kappa_k \langle\!\langle S_3^{(j)} S_3^{(k)} \rangle\!\rangle \Big]^{1/2},
\eeq
where we have used the cumulant notation $\langle\!\langle a b \rangle\!\rangle := \langle (a - \langle a \rangle)(b - \langle b \rangle) \rangle$.
As discussed in the previous section, the temperature is typically high compared to the dipole-dipole interaction energy. We can therefore ignore correlations of different spins and consider only the $j=k$ contribution, giving
\beq
\delta W_\mathrm{eff}^{FQ} &\approx& \cos(\theta) \Big[ \sum_j \kappa_j^2 \langle\!\langle (S_3^{(j)})^2 \rangle\!\rangle \Big]^{1/2}. 
\eeq
Considering a Nitrogen-to-NV conversion of $\eta\sim 5\%$, the sum over the impurities in the ensemble will be dominated by the Nitrogen impurities. An estimate of this expression can be obtained by noting that the coupling to the Nitrogen spin $\kappa$ is comparable to the coupling $g$ to the NVs. The difference between this sum and the sum leading to the collective coupling constant $G$ only differ by the number of terms in the sum and we obtain 
\beq \label{deltaWeff}
\delta W_\mathrm{eff}^{FQ} \sim \frac{\cos(\theta)}{\sqrt{\eta}} G.
\eeq
A more accurate treatment taking into account the full spatial distribution of the field over a sample of the dimensions considered in the text only changes the estimate by a factor of less than 2. Eq. (\ref{deltaWeff}) quantifies how close to the degeneracy point we need to be so that the dephasing of the FQ induced by the bath of spins is negligible. In particular, from (\ref{deltaWeff}) we derive that we can safely neglect this dephasing on a time scale $1/G$ for a conversion efficiency $\eta=0.05$ if $\cos(\theta)\lesssim 0.01 $, for which $\delta W_\mathrm{eff}^{FQ}/G \lesssim 0.05$.

Next we turn to the rapidly oscillating part of the coupling, described by the term containing $\hat{\tau}_1$, and assume that we are near the degeneracy point, that is $\sin \theta\approx1$. We will now argue that since $\hat{\tau}_1$ is oscillating rapidly, the slowly-varying dynamics of the spin bath will only have a very weak influence on the FQ. Specifically the spin bath can influence the FQ either through direct transitions to the spin bath or through a slow dephasing. The direct coupling can be excluded by noting that the coupling constant to collective excitations of the bath is limited by $G/\sqrt{\eta}$, but for reasonably low applied magnetic fields, $B^{{\rm ext}}\lesssim 10$ mT, the detuning $\omega-\Delta_j$ will be of the order of the NV zero-field splitting (a few GHz). The probability to transfer the excitation is $\sim G^2/(\omega-\Delta_j)^2\eta$, and can thus safely be neglected since $G\sim$ MHz. The dephasing of the FQ caused by the bath can be estimated by calculating the energy shift of the FQ in second order perturbation theory. Assuming that apart from the free precession, the spins change slowly on a time scale set by 
$\omega-\Delta_j$, the effective interaction is given by 
\begin{equation}
H_{{\rm eff}}=\hat \tau_3\sum_l\frac{{W}^{\rm FQ}_z (\vec{r}_l)^2}{4\omega}+\frac{{W}^{\rm FQ}_\perp (\vec{r}_l)^2\omega}{4(\omega^2-\Delta_j^2)}-\frac{{W}^{\rm FQ}_\perp (\vec{r}_l)^2\Delta_j}{2(\omega^2-\Delta_j^2)} S_z^{(j)}(t).
\end{equation}
Here the first two terms are independent of the state of the impurities, and therefore merely lead to a mean shift of the energy, which can be compensated by a magnetic field.  It is thus only the last term which leads to dephasing. Again, we can estimate the root mean square value of this term by neglecting the correlations among the impurities and we find a typical energy shift 
\begin{equation}
\Delta E \sim \frac{G}{\sqrt{\eta}} \frac{g\Delta_j}{\omega^2-\Delta_j^2},
\end{equation}
which we can safely neglect since the single-spin coupling $g\sim $ kHz is much smaller than any other quantities in the system. 

A remaining problem is the influence of the NVs with different orientations. The interaction with these centers is similar to that with the resonant subensemble and of order $\sim G$. We assume that we select a single orientation by applying a magnetic field with a component along the axis of the NV center. This will shift all other centers out of resonance by an amount $\delta\omega\sim g_e\mu_B B^{{\rm ext}}$, such that the resulting error can be estimated to be $G^2/\delta\omega^2$. For a reasonable applied field, e.g., $B^{{\rm ext}}\sim  10$ mT, this error is negligible.